\documentclass[reprint,twocolumn,superscriptaddress,amsmath,amssymb,showpacs,prl]{revtex4-1}
\usepackage{hyperref}

\usepackage{dcolumn}
\usepackage{epsfig}
\usepackage{graphics}
\usepackage[latin1]{inputenc} 
\usepackage[T1]{fontenc}
\usepackage{bm}
\newcommand{\ddst}{false}

\begin{document}

\title{Topological Control on Atomic Networks' Relaxation Under Stress}

 \author{Mathieu Bauchy}
 \email[Contact: ]{bauchy@ucla.edu}
 \homepage[\\Homepage: ]{http://mathieu.bauchy.com}
 \affiliation{Physics of AmoRphous and Inorganic Solids Laboratory (PARISlab), Department of Civil and Environmental Engineering, University of California, Los Angeles, CA 90095, United States}
 \author{Mengyi Wang}
 \affiliation{Physics of AmoRphous and Inorganic Solids Laboratory (PARISlab), Department of Civil and Environmental Engineering, University of California, Los Angeles, CA 90095, United States}
 \author{Yingtian Yu}
 \affiliation{Physics of AmoRphous and Inorganic Solids Laboratory (PARISlab), Department of Civil and Environmental Engineering, University of California, Los Angeles, CA 90095, United States}
 \author{Bu Wang}
 \affiliation{Physics of AmoRphous and Inorganic Solids Laboratory (PARISlab), Department of Civil and Environmental Engineering, University of California, Los Angeles, CA 90095, United States}
 \author{N M Anoop Krishnan}
 \affiliation{Physics of AmoRphous and Inorganic Solids Laboratory (PARISlab), Department of Civil and Environmental Engineering, University of California, Los Angeles, CA 90095, United States}
 \author{Franz-Joseph Ulm}
 \affiliation{Concrete Sustainability Hub, Department of Civil and Environmental Engineering, Massachusetts Institute of Technology, 77 Massachusetts Avenue, Cambridge, MA 02139, United States}
 \affiliation{MIT-CNRS joint laboratory at Massachusetts Institute of Technology, 77 Massachusetts Avenue, Cambridge, MA 02139, United States}
 \author{Roland Pellenq}
 \affiliation{Concrete Sustainability Hub, Department of Civil and Environmental Engineering, Massachusetts Institute of Technology, 77 Massachusetts Avenue, Cambridge, MA 02139, United States}
 \affiliation{MIT-CNRS joint laboratory at Massachusetts Institute of Technology, 77 Massachusetts Avenue, Cambridge, MA 02139, United States}
 \affiliation{Centre Interdisciplinaire des Nanosciences de Marseille, CNRS and Aix-Marseille University, Campus de Luminy, Marseille, 13288 Cedex 09, France}

\date{\today}


\begin{abstract}
Upon loading, atomic networks can feature delayed viscoplastic relaxation. However, the effect of composition and structure on such a relaxation remains poorly understood. Herein, relying on accelerated molecular dynamics simulations and topological constraint theory, we investigate the relationship between atomic topology and stress-induced relaxation, by taking the example of creep deformations in calcium--silicate--hydrates, the binding phase of concrete. Under constant shear stress, C--S--H is found to feature delayed logarithmic shear deformations. We demonstrate that the propensity for relaxation is minimum for isostatic atomic networks, which are characterized by the simultaneous absence of floppy internal modes of relaxation and eigen stress. This suggests that topological nano-engineering could lead to the discovery of non-aging materials.
\end{abstract}

\maketitle

\section{Introduction}

Out-of-equilibrium systems -- e.g., quenched glasses or jammed granular materials -- tend to relax over time towards a stable or meta-stable equilibrium state. In terms of energy landscape, such relaxation can be described as a succession of "jumps" between energy basins (local energy minima) through channels (modes of relaxation) \cite{mauro_continuously_2007, lacks_energy_2001, lacks_energy_2004}. Starting from a stable equilibrium state, external stress can deform the energy landscape, place the system in an out-of-equilibrium state, and, thereby, induce relaxation \cite{rottler_deformation_2008, bonn_laponite:_2002, viasnoff_rejuvenation_2002, utz_atomistic_2000, priezjev_heterogeneous_2013, lacks_energy_2001, lacks_energy_2004, fiocco_oscillatory_2013, lyulin_time_2007}. The temperature and height of the energy barriers then define the kinetics of relaxation \cite{vineyard_frequency_1957}.

Relaxation can result in delayed variations of volume or shape. This behavior is exemplified by creep, i.e., the delayed viscoplastic strain shown by a material under constant load. Although creep can affect, among others, metals, ceramics, or minerals \cite{poirier_creep_1985, barnes_yield_1999}, it is especially pronounced in concrete, even at ambient temperature, and can lead to the failure of structures \cite{bazant_prediction_2001, bazant_progress_2013, bazant_excessive_2011}. On the other hand, glasses, archetypical out-of-equilibrium systems, can feature long-term volume relaxation after being quenched \cite{welch_dynamics_2013, vannoni_long-term_2010, ruta_revealing_2014, sahu_room_2009}, a behavior known as the "thermometer effect" \cite{kurkjian_perspectives_1998, bunde_ionic_1998}.

Although the role of the composition and structure of atomic networks in controlling the propensity for relaxation remains poorly understood, specific glass compositions have been reported to feature little, if any, relaxation over time after quenching, which has been explained within the framework of topological constraint theory (TCT) \cite{phillips_topology_1979, phillips_topology_1981, thorpe_continuous_1983, mauro_topological_2011, bauchy_topological_2012}. Following Maxwell's study on the stability of mechanical trusses \cite{maxwell_l._1864}, TCT describes the rigidity of atomic networks, which can feature three distinct states: (1) \textit{flexible}, having internal degrees of freedom called floppy modes \cite{naumis_energy_2005} that allow for local deformations, (2) \textit{stressed--rigid}, being locked by their high connectivity, and (3) \textit{isostatic}, the optimal intermediate state (see Fig. \ref{fig:rigidity}a). The isostatic state is achieved when the number of constraints per atom, $n_c$, comprising radial bond-stretching (BS) and angular bond-bending (BB), equals three, the number of degrees of freedom per atom. Compositions characterized by an isostatic network have been found to exist inside a window \cite{feng_direct_1997}, located between the flexible ($n_c<3$) and the stressed--rigid ($n_c>3$) compositions, known as the Boolchand intermediate phase, and show some remarkable properties such as a stress-free character \cite{chubynsky_self-organization_2006}, a space-filling tendency \cite{rompicharla_abrupt_2008}, anomalous dynamical and structural signatures \cite{bauchy_transport_2013, bauchy_compositional_2013, micoulaut_anomalies_2013, bauchy_percolative_2013}, and maximum resistance to fracture \cite{bauchy_topological_2014}. Interestingly, isostatic networks have been shown to feature limited relaxation phenomena \cite{chakravarty_ageing_2005}.

Herein, relying on accelerated molecular dynamics (MD) simulations and TCT, we investigate the creep deformations under constant shear stress of calcium--silicate--hydrates (C--S--H), the phase that binds and control the properties of concrete, including creep \cite{taylor_cement_1997}. We show that, in analogy with glass relaxation, isostatic C--S--H compositions feature a low propensity for relaxation. In contrast, flexible and stressed-rigid networks show significant creep deformations, on account of the presence of low energy floppy modes of deformation and eigen stress, respectively.

\section{Simulation details}

In the present work, we rely on the C--S--H models developed by Pellenq \textit{et al.} \cite{pellenq_realistic_2009, abdolhosseini_qomi_combinatorial_2014}. The atomic models of C--S--H, with various compositions (different Ca/Si molar ratios), were obtained by introducing defects in an 11 \AA\ tobermorite configuration \cite{hamid_crystal-structure_1981} following a combinatorial approach \cite{abdolhosseini_qomi_combinatorial_2014}. This initial crystal consists of pseudo-octahedral calcium oxide sheets, surrounded on each side by silicate chains. These negatively charged calcium--silicate layers are separated from each other by both dissociated and undissociated interlayer water molecules and charge-balancing calcium cations. Starting from this structure, the Ca/Si ratio is gradually increased from 1.0 to 1.9 by randomly removing SiO$_2$ groups. The introduced defects offer possible sites for the adsorption of extra water molecules, which was performed via the Grand Canonical Monte Carlo method, ensuring equilibrium with bulk water at constant volume and room temperature. Eventually, the ReaxFF potential \cite{manzano_confined_2012}, a reactive potential, was used to account for the chemical reaction of the interlayer water with the defective calcium--silicate sheets. The use of a reactive potential allows us to observe the dissociation of water molecules into hydroxyl groups. The details of the methodology used for the preparation of the models, as well as multiple validations with respect to experimental data can be found in Ref. \cite{abdolhosseini_qomi_combinatorial_2014} and in previous works \cite{bauchy_order_2014,pellenq_realistic_2009,abdolhosseini_qomi_applying_2013,abdolhosseini_qomi_anomalous_2014,bauchy_nanoscale_2014,bauchy_fracture_2014,bauchy_is_2014,bauchy_topological_2014, bauchy_rigidity_2015}. In particular, this model has been shown to offer an excellent agreement with nano-indentation measurements of modulus and hardness \cite{vandamme_nanogranular_2010}, which renders it attractive to study creep. In this study, we keep the original ReaxFF potential, using a timestep of 0.25 fs. The samples were systematically relaxed to zero stress at 300 K before any further characterization.

We now focus on the methodology used to simulate creep. Traditional MD simulations are usually limited to a few nanoseconds, which prevents one from using them to predict long-term relaxation at low temperature. On the other hand, kinetic Monte-Carlo simulations \cite{barkema_event-based_1996} offer an attractive alternative to perform simulations up to a few seconds, but their application to silicate hydrates is challenging, e.g., due to the high mobility of the water molecules, which results in a huge number of small energy barriers to compute. Since a direct simulation of the stress-induced relaxation dynamics of C--S--H is, at this point, unachievable, we applied a method that has recently been introduced to study the relaxation of silicate glasses \cite{yu_stretched_2015}. In this method, the system is subjected to small, cyclic perturbations of shear stress $\pm \Delta \tau$ around zero pressure. At each stress cycle, a minimization of the energy is performed, with the system having the ability to deform (shape and volume) in order to reach the target stress. Note that the observed relaxation does not depend on the choice of $\Delta  \tau$, provided that this stress remains sub-yield \cite{yu_stretched_2015}.

This method mimicks the artificial aging observed in granular materials subjected to vibrations \cite{richard_slow_2005, mobius_irreversibility_2014}. Indeed, small vibrations induce a compaction of the material, that is, they make the system artificially age. On the other hand, large vibrations randomize the grain arrangements, which decreases the overall compactness and, therefore, make the system rejuvenate. Similar ideas, relying on the energy landscape approach \cite{lacks_energy_2001, lacks_energy_2004}, have been applied to amorphous solids, based on the fact that small stresses deform the energy landscape undergone by the atoms. This can result in the removal of some energy barriers existing at zero stress, thus allowing atoms to jump over them in order to relax to lower energy states. This transformation is irreversible as, once the stress is removed, the system remains in its "aged" state. In contrast, large stresses move the system far from its initial state, which eventually leads to rejuvenation, similar to thermal annealing \cite{utz_atomistic_2000, lyulin_time_2007}. Here, in order to mimick deviatoric creep deformation, we add to the previous method a constant shear stress $\tau_0$, such that $\Delta \tau < \tau_0$ (see the inset of Fig. \ref{fig:method}).

\section{Results}

\begin{figure}
	\begin{center}
		\includegraphics*[width=0.7\linewidth, keepaspectratio=true, draft=\ddst]{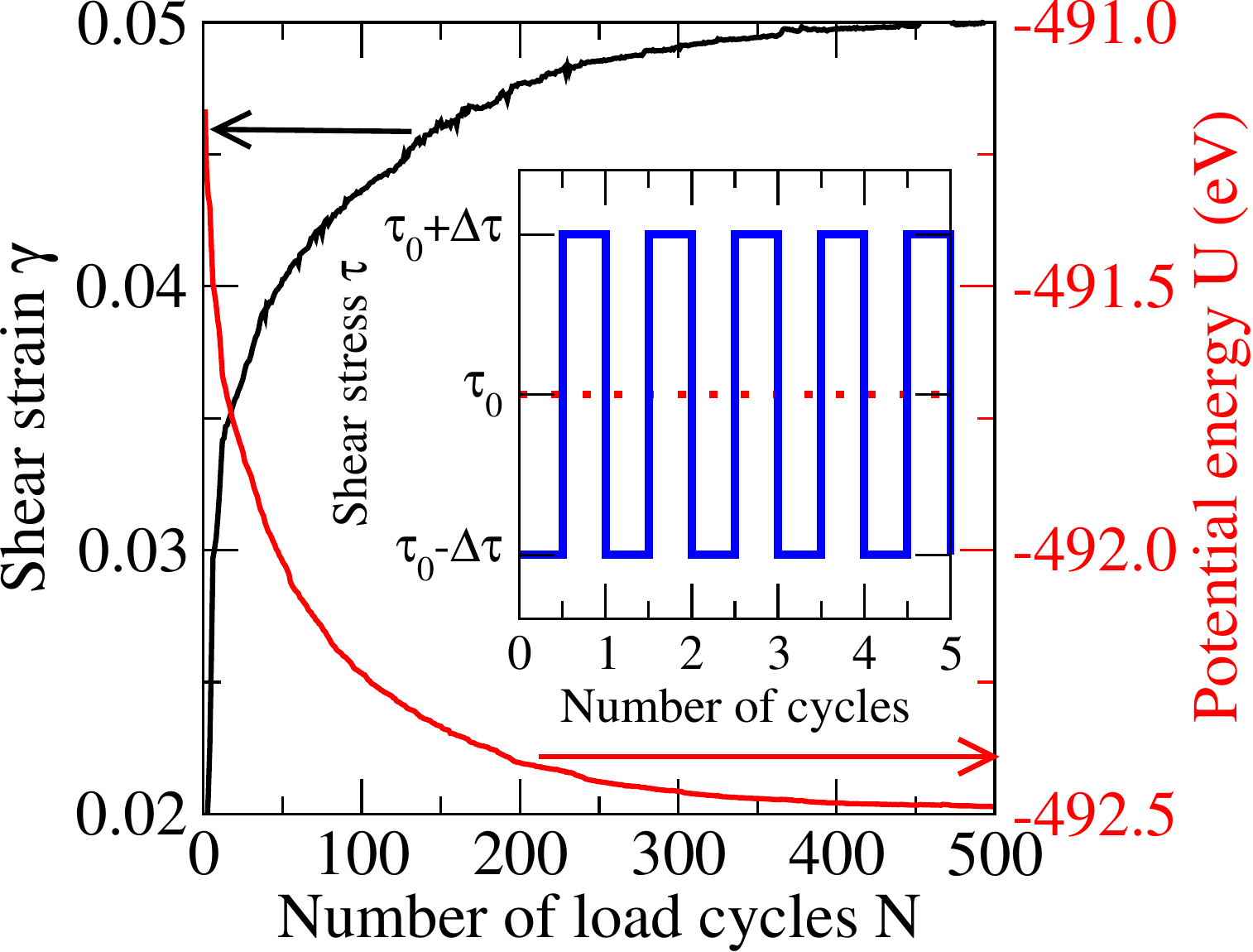}
		\caption{\label{fig:method} Shear strain (left axis) and potential energy (right axis) with respect to the number of loading/unloading cycles. The inset shows the shape of the applied shear stress.
		}
	\end{center}	
\end{figure}

\begin{figure}
\begin{center}
\includegraphics*[width=\linewidth, keepaspectratio=true, draft=\ddst]{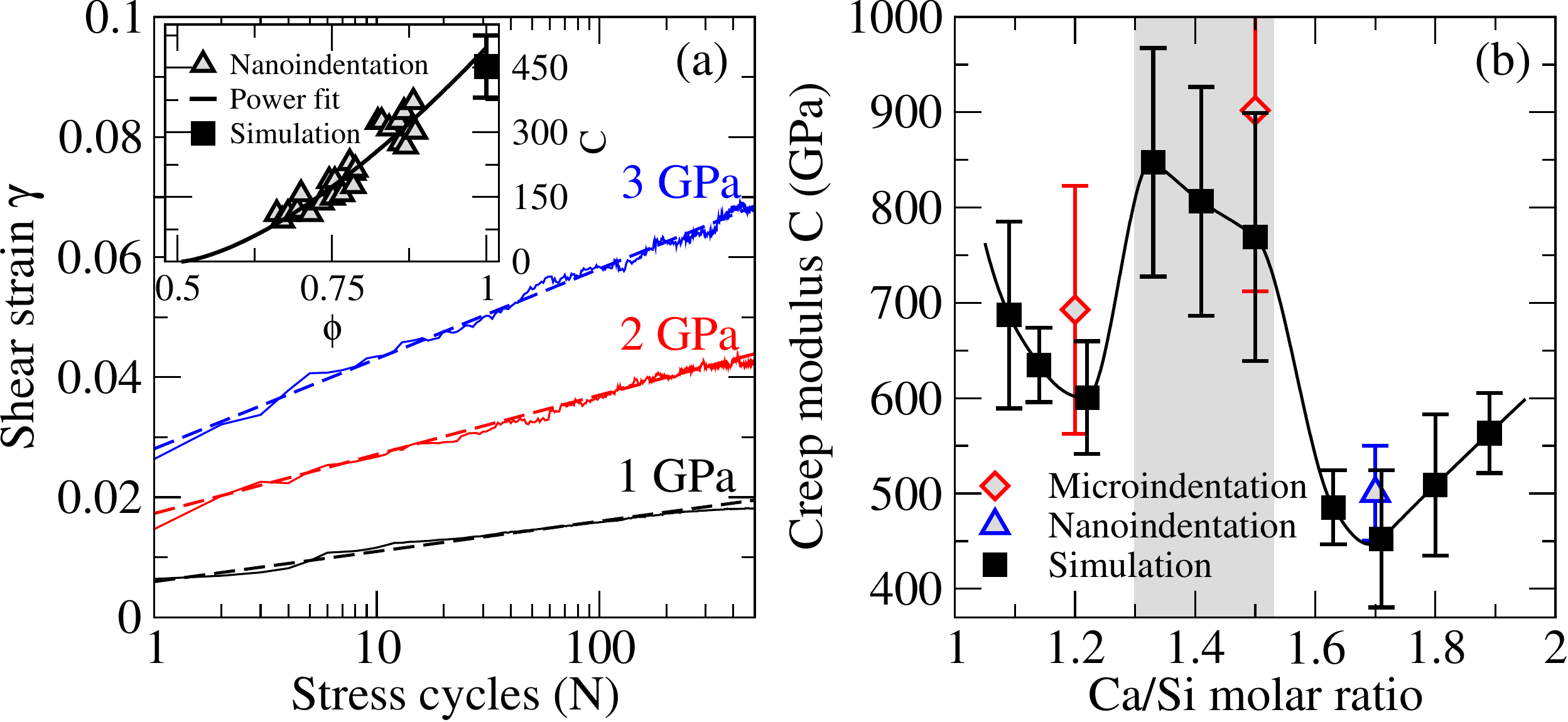}
\caption{\label{fig:strain} (a) Computed shear strain $\gamma$, for Ca/Si = 1.7, with respect to the number of loading/unloading cycles $N$, under a constant shear stress $\tau_0$ of 1, 2, and 3 GPa. The dashed lines indicate logarithmic fits following $\gamma = (\tau_0/C) \log (1 + N/N_0)$, permitting to evaluate the creep modulus $C$. The inset shows the creep modulus $C$ with respect to the packing fraction $\phi$ obtained from nanoindentation \cite{vandamme_nanogranular_2010}. The values are fitted by a power law $C = A (\phi - 0.5)^{\alpha}$ and extrapolated to $\phi = 1$ to be compared with the value obtained by the present simulations. (b) Computed creep modulus $C$ with respect to the Ca/Si molar ratio. The values are compared with experimental measurements obtained by micro-indentation \cite{nguyen_microindentation_2014} and nano-indentation \cite{vandamme_nanogranular_2009}. The grey area indicates the extent of the compositional window in which a maximum resistance to creep is observed.
}
\end{center}	
\end{figure}

As shown in Fig. \ref{fig:method}, the application of such stress cycles results in the gradual increase and decrease of the shear strain and potential energy, respectively. This confirms that, upon creep, C--S--H relaxes towards lower energy states. As shown in Fig. \ref{fig:strain}a, we observe that, when subjected to shear stresses $\tau_0$ of different intensities, C--S--H presents a shear strain $\gamma$ that increases logarithmically with the number of cycles $N$. Such a logarithmic trend is in agreement with experimental observations \cite{vandamme_nanogranular_2010}, which suggests that the creep deformation can be expressed as:

\begin{equation}
\gamma (N) = (\tau_0/C) \log (1 + N/N_0)
\end{equation} where, $N_0$ is a fitting parameter analogous to a relaxation time and $C$ is the creep modulus, which is the inverse of the creep compliance $S$ and can be determined by fitting the computed shear strain with respect to the number of stress cycles (see Fig. \ref{fig:strain}a). Careful analysis of the internal energy shows that the height of the energy barriers, through which the system transits across each cycle, remains roughly constant over successive cycles. On the basis of transition state theory, which specifies that the time needed for a system to jump over an energy barrier $E_{\rm a}$ is proportional to $\exp(-E_{\rm a}/kT)$, one can assume that each cycle corresponds to a constant duration $\Delta t$, so that a fictitious time can be defined as $t = N \Delta t$ \cite{masoero_kinetic_2013}. Note that these transitions would spontaneously occur, although the duration $\Delta t$ before each jump is too long to be accessible from conventional atomistic simulations.

Interestingly, we find that the computed shear strains are proportional to the applied constant shear stress $\tau_0$. As such, $C$ does not depend on the applied stress and, thereby, appears to be an intrinsic property of the material. We observe, however, that this holds only as long as the applied stress remains lower than the yield stress of the sample \cite{bauchy_creep_2015}. Note that, as our simulation do not consider any porosity, the computed values of C can only be compared with experimental values extrapolated to zero porosity. As shown in the inset of Fig. \ref{fig:strain}a, the obtained $C$ ($\sim 450$ GPa) is in very good agreement with nano-indentation data \cite{vandamme_nanogranular_2010}, extrapolated to a packing fraction of 1. To the best of our knowledge, this is the first time that the creep propensity (indicated by the creep modulus) of cementitious, or other viscoelastic materials has been successfully reproduced by atomistic simulation.

Further, to better understand the relationship between composition and stress relaxation propensity, the same approach was used for other C--S--H compositions. As shown in Fig. \ref{fig:strain}b, the computed $C$ values show a non-linear evolution with Ca/Si, which manifests in the form of a broad maximum around Ca/Si = 1.5, with a creep modulus around 80 \% higher than that obtained for Ca/Si = 1.7. Such a non-linear behavior is very different from those of indentation hardness and creep modulus, as both of them decrease monotonically with Ca/Si \cite{abdolhosseini_qomi_combinatorial_2014}. Once again, the obtained $C$ values are in excellent agreement with micro-indentation data extrapolated to zero porosity \cite{nguyen_microindentation_2014}, which strongly suggests that the present method offers a realistic description of the creep of C--S--H at the atomic scale.

\section{Dicussion}

\begin{figure}
	\begin{center}
		\includegraphics*[width=\linewidth, keepaspectratio=true, draft=\ddst]{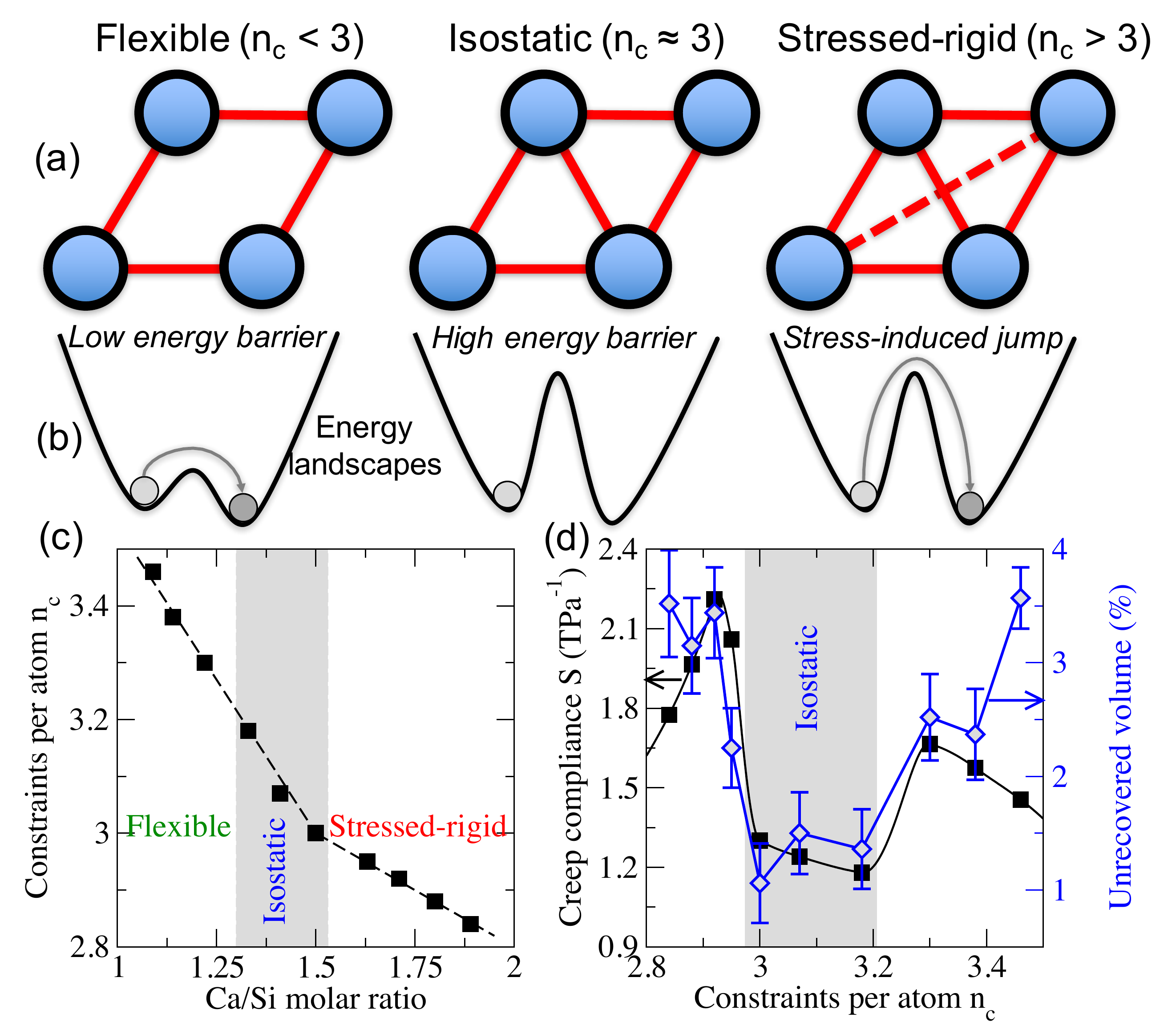}
		\caption{\label{fig:rigidity} (a) The three rigidity states of an atomic network. (b) Schematics of the corresponding energy landscapes. (c) Number of topological constraints per atom $n_{\rm c}$ in C--S--H with respect to Ca/Si \cite{bauchy_rigidity_2015}. The dashed line is a guide for the eye. The grey area indicates the extent of the compositional window in which a maximum resistance to creep is observed (see Fig. \ref{fig:strain}b), also corresponding to the range of isostatic compositions ($n_{\rm c} \simeq 3$), effectively separating the flexible ($n_{\rm c} < 3$) from the stressed-rigid domains ($n_{\rm c} > 3$). (d) Creep compliance $S$, the inverse of the creep modulus $C$, and fraction of unrecovered volume after loading and unloading with respect to $n_{\rm c}$. The grey area indicates the extent of the isostatic compositional window (see above). The line is a guide for the eye.			
		}
	\end{center}	
\end{figure}

We now investigate how the atomic topology controls the propensity for creep relaxation. As shown in Fig. \ref{fig:rigidity}c, C--S--H has been reported to feature a composition-induced rigidity transition at Ca/Si = 1.5 \cite{bauchy_rigidity_2015}, being stressed-rigid ($n_{\rm c} > 3$) at lower Ca/Si and flexible ($n_{\rm c} < 3$) at higher Ca/Si. As such, as shown in Fig. \ref{fig:rigidity}d, isostatic compositions ($n_{\rm c} \simeq 3$) feature the lowest propensity for creep, that is, the lowest creep compliance. To the best of our knowledge, these results constitute the first quantitative evidence of a link between atomic topology (hence composition) and resistance to stress relaxation, and suggest that the isostatic compositional window (1.3 < Ca/Si < 1.53) is analogous to a rigid but free of eigen stress Boolchand intermediate phase \cite{rompicharla_abrupt_2008}.

Interestingly, isostatic glassy networks have been shown to behave largely reversibly with stress, that is, to show nearly complete elastic recovery after compression \cite{varshneya_microhardness_2007, mauro_modeling_2007}. As such, to further demonstrate the analogy between isostatic C--S--H composition and glasses belonging to a Boolchand intermediate phase, the C--S--H samples were hydrostatically compressed under 10 GPa during 1 ns and subsequently relaxed at zero pressure to assess the extent of loading-induced permanent densification. As shown in Fig. \ref{fig:rigidity}d, we observe that isostatic C--S--H compositions indeed show the lowest unrecovered volume after loading/unloading. This feature can be explained as follows. (1) Thanks to their internal floppy modes, flexible systems can easily undergo irreversible deformations during loading. This enables irreversible structural deformations upon loading. (2) In contrast, stressed-rigid systems are completely locked. Once compressed, the high connectivity prevents the full relaxation of the accumulated internal stress, so that the network remains permanently densified after unloading. (3) Eventually, isostatic systems, i.e., rigid but free of eigen stress, simply adapt with pressure in a reversible way.

Our results can be understood within the energy landscape framework. The energy landscape of an atomic network is determined by the densities of bond and floppy mode, wherein the bond density tend to induce the creation of energy basins, whereas the floppy mode density leads to the formation of channels between the basins. (1) On account of their internal floppy modes, flexible atomic networks feature some low energy modes of internal reorganization to relax any loading-induced internal stress. Such floppy modes extend the number of energy channels, thereby enhancing the propensity for creep. (2) In contrast, stressed-rigid atomic networks show some internal eigen stress \cite{chubynsky_self-organization_2006} as all constraints cannot be simultaneously satisfied. This eigen stress induces an instability of the network and, therefore, acts as a driving force for phase separation or devitrification in glasses \cite{mauro_topological_2011, micoulaut_constrained_2008}. Such a driving force facilitates jumps between the basins, which, again, extends the possibilities of creep relaxation. In addition, the simulation method of creep implemented herein suggests that creep can be seen as a succession of small cycles of stress. Hence, the fact that both flexible and stressed-rigid networks feature low elastic recovery after loading explains their gradual deformation during creep. (3) Finally, isostatic networks, which are free of both internal modes of deformation and eigen stress, simultaneously do not feature any barrier-less channel between energy basins or eigen stress-induced driving force for relaxation. As such, such optimally constrained networks feature the lowest propensity for stress relaxation and creep.

\section{Conclusion}

Altogether, these results highlight the strong relationship between atomic topology and propensity for relaxation. Beyond concrete creep, being able to understand, predict, and control the relaxation and aging of materials could improve the understanding of memory encoded materials \cite{fiocco_encoding_2014, fiocco_memory_2015} or protein folding \cite{mousseau_sampling_2001, phillips_scaling_2009}. This also suggests that topological nano-engineering is a valuable tool to explore new compositional spaces, for the discovery of new materials featuring unusual properties.

\begin{acknowledgments}
	This material is based upon work supported by the National Science Foundation under Grant No. 1562066. This work was also supported by Schlumberger under an MIT-Schlumberger research collaboration and by the CSHub at MIT. This work has been partially carried out within the framework of the ICoME2 Labex (ANR-11-LABX-0053) and the A*MIDEX projects (ANR-11-IDEX-0001-02) cofunded by the French program "Investissements d'Avenir" which is managed by the ANR, the French National Research Agency.
\end{acknowledgments}

\end{document}